\documentclass[upint,varvw,barcolor=black,mathalfa=cal=euler,balance,hyphenate,french,pdf-a]{asmejour}% subscriptcorrection,
\usepackage{booktabs}
\hypersetup{%
	pdfauthor={Varun Kulkarni},          	% <=== change to YOUR name[s]!
	pdftitle={Liquid Sheet Breakup in Gas-Centered Swirl Coaxial Atomizers},    % <=== change to YOUR pdf file title
	pdfkeywords={gas-centered coaxial atomizers, liquid sheets, jet/sheet breakup, sprays},% <=== change to YOUR pdf keywords
	pdfsubject = {Describes the asmejour LaTeX template},			% <=== change to YOUR subject
	pdflicenseurl={https://ctan.org/pkg/asmejour},% may delete
}
\JourName{Fluids Engineering}%<=== change to the name of your journal

\begin{document}

% Change to your author name[s] and addresses, in the desired order of authors.
% First name, middle initial, last name
% Use title case (upper and lower case letters)
% Note usage below for corresponding author.

\SetAuthorBlock{V. Kulkarni \\ \vspace{5pt} D. Sivakumar\CorrespondingAuthor \\ \vspace{7pt} C. Oommen}{\vspace{5pt} Department of Aerospace Engineering,\\
   Indian Institute of Science,\\
   Bangalore,\\
   Karnataka 560012, India\\
   email: dskumar@aero.iisc.ac.in} 
% To label one or more corresponding authors put "Name\CorrespondingAuthor". No space after "Name".
% An optional argument can be added if email is not in address block as
%      "Name\CorrespondingAuthor{write@to.me}"
% Can also include multiple emails and use the command more than once for multiple corresponding authors,
%      "Name\CorrespondingAuthor{write@to.him, write@to.her}"
\SetAuthorBlock{T. J. Tharakan}{\vspace{5pt}
Liquid Propulsion Systems Centre,\\
Indian Space Research Organization,\\
Valiamala,\\
Thiruvananthapuram 695547, India\\
}

%%% Change to your paper title. Can insert line breaks if you wish (otherwise breaks are selected automatically).
\title{Liquid Sheet Breakup in Gas-Centered Swirl Coaxial Atomizers}
%%% Change these to your keywords.  Keywords are automatically printed at the end of the abstract.
%%% This command must come BEFORE the end of the abstract.
%%% If you don't want keywords, omit the \keyword{..} command.

%% Abstract should be no more than 250 words
\begin{abstract}
The study deals with the breakup behavior of swirling liquid sheets discharging from gas-centered swirl coaxial atomizers with attention focused toward the understanding of the role of central gas jet on the liquid sheet breakup. Cold flow experiments on the liquid sheet breakup were carried out by employing custom fabricated gas-centered swirl coaxial atomizers using water and air as experimental fluids. Photographic techniques were employed to capture the flow behavior of liquid sheets at different flow conditions. Quantitative variation on the breakup length of the liquid sheet and spray width were obtained from the measurements deduced from the images of liquid sheets. The sheet breakup process is significantly influenced by the central air jet. It is observed that low inertia liquid sheets are more vulnerable to the presence of the central air jet and develop shorter breakup lengths at smaller values of the air jet Reynolds number \textit{Re}\textsubscript{g}. High inertia liquid sheets ignore the presence of the central air jet at smaller values of \textit{Re}\textsubscript{g} and eventually develop shorter breakup lengths at higher values of \textit{Re}\textsubscript{g}. The experimental evidences suggest that the central air jet causes corrugations on the liquid sheet surface, which may be promoting the production of thick liquid ligaments from the sheet surface. The level of surface corrugations on the liquid sheet increases with increasing \textit{Re}\textsubscript{g}. Qualitative analysis of experimental observations reveals that the entrainment process of air established between the inner surface of the liquid sheet and the central air jet is the primary trigger for the sheet breakup. \href{https://doi.org/10.1115/1.4000737}{\textup{[DOI: 10.1115/1.4000737]}}
\end{abstract}

%\date{Version \versionno, \today}%% You can modify this information as desired. 
\date{}
							%% Putting \date{} will suppress any date.  
							%% If this command is omitted, date defaults to \today
							%% This command must come somewhere before \maketitle

\maketitle %% This command creates the author/title/abstract block. Essential!

%%%%%%%%%%%%%%%%%%%%%%%%%%%%%%%%%%%%%%%%%%%%%%%%%%%%%%%%%%%%%%%%%%%%%%%%%%%%%%%%%%%%%%%%%%%%%%%%%%%%%%%
%%%%%%%%%%%%%%%%%%%%%  End of fields to be completed. Now write! %%%%%%%%%%%%%%%%%%%%%%%%%%%%%%%%%%%%%%

\section{Introduction}

\begin{figure}
\centering\includegraphics[scale=0.8]{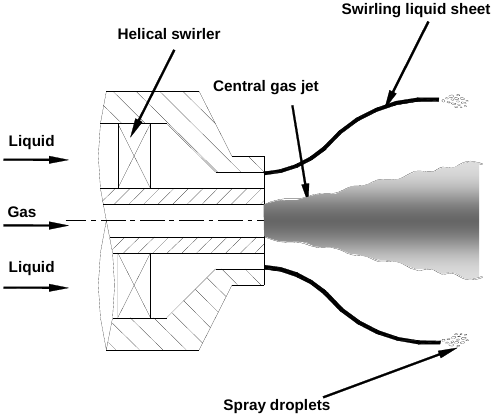}
\caption{\label{Fig1} A schematic of jets discharging from a gas-centered swirl coaxial atomizer}
\end{figure}

It has always been the endeavor of all engineers working in spray combustion to transform bulk liquid fuel and/or oxidizer into tiny droplets. A decrease in the size of liquid droplets results in better combustion quality parameters such as combustion efficiency, heat release, exhaust gas emissions, etc. Spray formation is accomplished in a liquid propellant engine by means of an atomizer present inside the engine combustion chamber. A successful atomizer design must provide high quality of atomization needed to ensure an efficient combustion process. For liquid propellant rocket LPR applications, several atomizer configurations are being currently employed and the most commonly used configurations are the impinging type \citep{Ryan1995, Ashgriz2001} and the coaxial type \citep{Mayer1994, Rahman1995, Sivakumar1996, Sivakumar1998, Cohn2003}. A coaxial type atomizer comprises of an inner orifice surrounded by an outer orifice and it has been used with both liquid/gas and liquid/liquid type propellant combinations. The present investigation of gas-centered coaxial swirl atomizer falls under the category of liquid/gas coaxial atomizers. Gas-centered swirl atomizers have been studied under two major configurations: the premixing type and the coaxial type. In the premixing type, the liquid flows over the inner wall of the gaseous orifice by passing it through tangential ports and a mixing process between the liquid and the gas occurs inside the gaseous orifice. In the coaxial type, the liquid and the gas flow via two different orifices arranged coaxially. Figure \ref{Fig1} illustrates a schematic of a coaxial type gas-centered swirl atomizer. The atomizer, as illustrated in the figure, discharges a central gaseous jet surrounded by an annular swirling liquid sheet. Soller et al. \citep{Soller2005} studied the combustion behavior of sprays discharging from gas-centered swirl coaxial atomizers and found that the coaxial type atomizers exhibit superior combustor wall compatibilities compared with that of the premixing type in the context of wall thermal stress levels. Sprays discharging from the coaxial type gas-centered swirl atomizers bend toward the spray axis, thereby reducing the thermal load experienced on the walls of combustor\citep{Muss2002}. 

Spray formation in atomizers is governed by the breakup behavior of liquid jet or sheet discharging from the atomizer orifices\citep{Lefebvre1989}. In gas-centered swirl coaxial atomizers, the physical process by which the outer liquid sheet interacts with the central gas jet and the role of the central gas jet in the breakup of the liquid sheet into drops are key factors in determining the quality of atomization. The fundamental mechanisms governing the breakup of a liquid sheet injected into still ambient air are well documented in the current literature. The comprehensive experimental study by Dombrowski and Fraser \citep{Dombrowski1954} revealed the influence of flow and orifice parameters such as liquid properties, velocity of the liquid sheet, turbulence at the orifice exit, etc., on the breakup of liquid sheets. The breakup occurs via the interaction of the liquid sheet with the surrounding gas medium \citep{Squire1953, Taylor1959a, Taylor1959b}. The aerodynamic shear developed over the surfaces of the liquid sheet causes waves to grow and these waves separate from the liquid sheet in the form of ligaments, which subsequently disintegrated into droplets. Two types of waves aid the sheet breakup: sinuous and dilational\citep{Hagerty1955}. Sinuous waves are produced when the two surfaces of the liquid sheet oscillate in phase with each other and dilational waves result from the out-of-phase motion of the two surfaces. Several theoretical studies have been reported on the propagation of these waves in liquid sheets and their influence on the sheet stability, and a recent review summarizes the outcomes of these investigations\citep{Lin2003}.

The influence of coflowing air on the breakup behavior of liquid sheets, particularly planar liquid sheets, has been studied extensively. Mansour and Chigier \citep{Mansour1990} studied the disintegration mechanism of liquid sheets discharging from a slit type orifice in the presence of coflowing air. The airflow over the surfaces of the liquid sheet augments the formation of cellular structures thin liquid membrane bounded by thick ligaments in the liquid sheet. In a similar study, Stapper et al. \citep{Stapper1992} suggested that the cellular type breakup occurs at higher relative velocities between the liquid sheet and the coflowing air jet. A much finer experimental study on the formation of cellular structures in liquid sheets by Park et al.\citep{Park2004} showed that their presence is attributed to the unstable sinuous waves developing over the liquid sheet. The cellular type breakup results in wider drop size histograms as the thin membranes of cellular structures generate finer spray droplets and the rims seen at the periphery of cellular structures generate larger diameter droplets. A classification of disintegration regimes for liquid sheets with low air velocities discharging from a coaxial airblast atomizer was done by Adzic et al. \citep{Adzic2001}. The authors identified three major disintegration regimes: Kelvin-Helmholtz regime, cellular regime, and atomization regime. A systematic study on the role of coflowing air on the liquid sheet behavior was carried out by Lozano et al.\citep{Lozano2001} by including the effect of air boundary layer and air viscosity in the stability analysis of the liquid sheet. The authors suggested that the air velocity mostly determines the oscillatory behavior of the liquid sheet and the theoretical predictions on the characteristics of sheet oscillation improve with the inclusion of air viscosity in the analysis. A general observation in these studies is that the flow characteristics of coflowing air play a dominant role in the disintegration of the liquid sheet sandwiched between the air sheets. 

The present experimental study deals with the breakup phenomena of swirling liquid sheets discharging from gas-centered swirl coaxial atomizers. The primary objective is to understand how the central gas jet aids the breakup process of such liquid sheets and the subsequent spray formation. The breakup of swirling liquid sheets without the coflowing gas jet has been studied extensively in the context of pressure swirl atomizers, and Lefebvre \citep{Lefebvre1989} provides a comprehensive summary on the breakup process of such simple swirling liquid sheets. In the present work, a gas jet exists in the air core of the abovementioned simple swirling liquid sheets. The effect of the central gas jet on the sheet behavior is significantly different from the role played by still ambient air, and such an interaction process between the central gas jet and the swirling liquid sheet is not fully understood in the current literature. Note that several of the published works on gas-centered swirl coaxial atomizers confined their attention to the analysis of combustion related aspects \citep{Cohn2003, Soller2005, Muss2002}. A systematic experimental investigation is carried out in the present work to document the influence of central gas jet on the breakup behavior of the outer swirling liquid sheet by conducting cold flow experiments at ambient atmospheric conditions with air and water as working fluids.

\section{Experimental Details}
\begin{figure}
\centering\includegraphics[width=\linewidth]{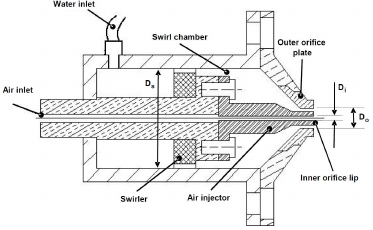}
\caption{\label{Fig2}A schematic of the gas-centered swirl coaxial atomizer used in the present study}
\end{figure}

A schematic of the gas-centered swirl coaxial atomizer used in this study is shown in Fig. \ref{Fig2}. Major components of the atomizer configuration are highlighted in the figure. The outer orifice plate consisted of a converging portion and an orifice. Diameter of the outer orifice $D_o$ and that of the inner central orifice $D_i$ were kept at 4.4 mm and 2.4 mm, respectively. Lip thickness of the inner orifice was kept at 0.3 mm, which resulted in an orifice gap of 0.7 mm for the outer orifice. Swirling motion to the flowing liquid was imparted by passing the liquid through rectangular shaped helical passages present over the periphery of the swirler. Two different atomizer configurations CA1 and CA2 were studied by varying the geometrical parameters of the swirler. The width $w$ of the rectangular helical passage was kept at 1.0 mm for both CA1 and CA2. The depth h of the helical passage was kept differently for the atomizer configurations as 0.5 mm for CA1 and 1.0 mm for CA2. The number of helical passages in the swirler $n$ was fixed at 6 for both the atomizer configurations. The swirling intensity of the outer pressure swirl atomizer was characterized in terms of swirl number $S$, which was estimated in the present study using the relation,
\begin{equation}
S = \dfrac{\pi D_sD_o}{4nwh}
\end{equation}
where, $D_s$ is the diameter of the swirl chamber. The estimated values of $S$ for the atomizer configurations CA1 and CA2 were
25.7 and 12.3, respectively. A standard spray test facility was used to carry out the experiments. The facility consisted of a large
stainless steel tank containing water kept under a particular pressure by means of a compressed air supply and an air pressure
regulator. Flexible tubes capable of withstanding high pressures were used to carry the experimental liquid to the atomizer assembly. Tubes similar in nature, drawn from the compressed air supply, were used to deliver air at a given mass flow rate which was estimated by employing an orifice flow meter. A filter was employed between the liquid storage tank and the atomizer assembly
to arrest contaminants present in the liquid. Necessary needle valves along with pressure gauges were connected in the flow
lines to keep the fluid flow at steady operational conditions. 

\begin{figure}
\centering\includegraphics[scale=1.2]{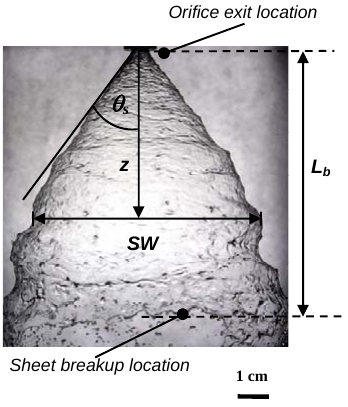}
\caption{\label{Fig3}A typical image of the outer liquid sheet illustrating the details of $\theta_s$, \textit{L}\textsubscript{b}, and \textit{SW}}
\end{figure}

The flow behavior of liquid sheets was characterized by capturing images of liquid sheets using photographic techniques. A Nikon D1X digital camera with diffused backlighting system was used to take photographs of the liquid sheets. The pixel resolution of the camera was 2000 $\times$ 1312. The diffused backlighting system consisted of a strobe lamp, and each flash from the strobe lamp had lasted for about 12 $\mu$s, which became the time resolution for imaging. The measurements obtained from the image analysis are: the half spray cone angle of spray at the orifice exit, the radial spread of spray $SW$ at different locations in the spray axis, the breakup length of the liquid sheet $L_b$, and the two-dimensional surface profile of the liquid sheet. Figure \ref{Fig3} shows the details of $L_b$, and $SW$ on a typical image of the outer swirling liquid sheet.
An image processing algorithm was developed to deduce the two dimensional surface profile of the liquid sheet in the form of plots
from the high resolution images of liquid sheets. The algorithm identifies the bounding pixels of spray contour at different $Z$ locations counted in terms of pixels from the orifice exit. 
The locations of the bounding pixels in the digital image were carefully extracted so that the data can be plotted using a graphical software. These profiles were analyzed systematically to understand the surface perturbations and ligaments developing from the liquid sheets. The flow condition of the outer liquid sheet was expressed in terms of Weber number We$_l$ as,
\begin{equation}
We_l = \dfrac{\rho_l U_l t_f}{\sigma}
\end{equation}
where, $\rho_l$ is the liquid density 998 kg/m\textsuperscript{3}, $\sigma$ is the surface tension (0.0728 N/m), U$_l$ is the axial velocity of the liquid sheet at the orifice exit, and $t_f$ is the thickness of the liquid sheet at the orifice exit. The value of $t_f$ was estimated using an analytical relation reported in the literature as \citep{Lefebvre1989},
\begin{equation}
t_f = 3.66\left(\dfrac{D_o m_l\mu_l}{\rho_l (\Delta P_l)}\right)^{0.25}
\end{equation}
where $m_l$ is the liquid mass flow rate, $P_l$ is the injection pressure drop across the outer orifice, and $\mu_l$ is the liquid dynamic viscosity (1.003$\times$10\textsuperscript{-3} kg/ms). The axial velocity of the liquid sheet at the orifice exit, U$_l$ was estimated from the mass conservation as,
\begin{equation}
U_l = \dfrac{m_l}{\rho_l \pi t_f(D_o - t_f)}
\end{equation}
The flow condition of the central air jet was expressed in terms of Reynolds number $Re_g$ as
\begin{equation}
Re_g = \dfrac{\rho_g U_g D_i}{\mu_g}
\end{equation}
where, $\rho_g$ is the air density (1.2 kg/m\textsuperscript{3}), $U_g$ is the axial velocity of the gas jet at the orifice exit, and $\mu_g$ is the air dynamic viscosity (1.81$\times$10\textsuperscript{-5} kg/ms). 
\begin{table}[t]
\newcolumntype{C}{>{$}c<{$}} % math-mode version of "c" column type, from array package
\caption{\label{Table1}Uncertainty estimates}
\small
\addtolength{\tabcolsep}{-0.6em}
\begin{tabular*}{\columnwidth}{ccc}
\toprule
\textrm{Quantity} & \textrm{Source} & \textrm{Net Uncertainty} \rule{0pt}{8pt}\\
\midrule
\textrm{$m_l$}  & \textrm{Measurement resolution (1\%)} & \textrm{1–4.15\%}\\
           & \textrm{Repeatability (0.12 - 4.03\%)} & \\[2pt]
\textrm{$m_g$}  & \textrm{Measurement resolution (0.030\%)} & \textrm{1.75–4.68\%}\\
           & \textrm{Repeatability (0.22–4.3\%)} & \\[2pt]
\textrm{$(\Delta P)_l$}  & \textrm{Measurement resolution (2\%)} & \textrm{2.6–4.28\%}\\
           & \textrm{Repeatability (1.67–3.57\%)} & \\[2pt]
           & \textrm{Calibration (0.13–1.27\%)} & \\[2pt]
\textrm{$(\Delta P)_g$}  & \textrm{Measurement resolution (0.02\%)} & \textrm{1.9–8.42\%}\\
           & \textrm{Repeatability (0.83–6.67\%)} & \\
           & \textrm{Calibration (1.71–5.16\%)} & \\[2pt]
\textrm{$L_b$}  & \textrm{Measurement resolution (1\%)} & \textrm{1.35–10.75\%}\\
           & \textrm{Repeatability (1.35–10.75\%)} & \\[2pt]
\textrm{$SW$}  & \textrm{Spray width point identification (0.53–2.31\%)} & \textrm{3.1–8.03\%}\\
           & \textrm{Repeatability (3.05–7.69\%)} & \\
\bottomrule
\end{tabular*}
\end{table}
The value of $U_g$ was estimated from the mass conservation as,
\begin{equation}
U_g = \dfrac{4m_g}{\rho_g \pi D_i^2}
\end{equation}
where, $m_g$ is the air mass flow rate. The uncertainty estimates for primary measurements are presented in Table \ref{Table1}. The uncertainty estimates for derived  measurements like spray contraction parameter, nondimensionalized breakup length, etc., can be determined from Table \ref{Table1}. The calculation for net uncertainty was based on the procedure described by Moffat\citep{Moffat1988}.

\section{Results and Discussion}
\subsection{Visual Observations on the Breakup of the Outer Liquid Sheet by the Central Air Jet}
\begin{figure}[b!]
\centering\includegraphics[width=\linewidth]{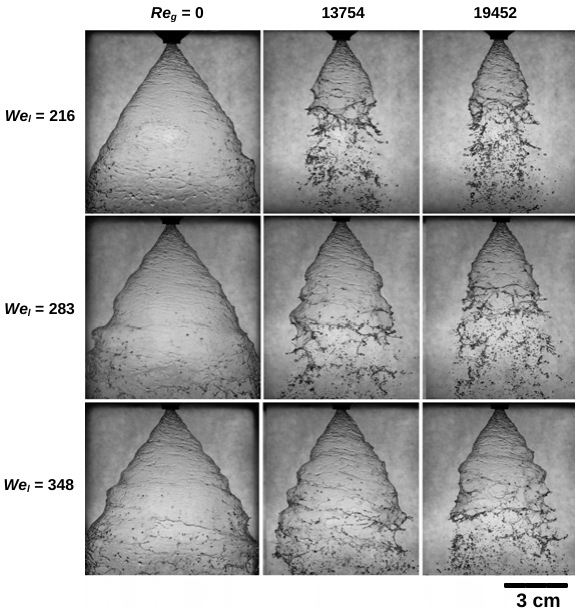}
\caption{\label{Fig4} Images of liquid sheets discharging from the gas centered swirl coaxial atomizer CA1 ($S$ = 25.7) for different combinations of We$_l$ and Re$_g$}
\end{figure}
Figure \ref{Fig4} illustrates the images of outer swirling liquid sheets discharging from the gas-centered swirl coaxial atomizer CA1 with different combinations of outer liquid sheet and central air jet flow conditions.
\begin{figure*}[t!]
\centering\includegraphics[scale=0.9]{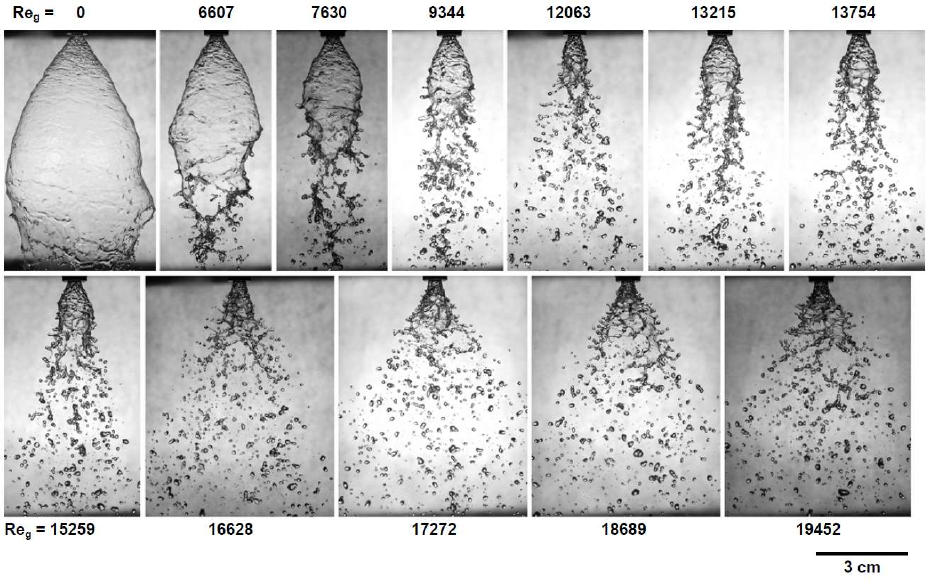}
\caption{\label{Fig5}The influence of the central air jet on the flow behavior of the outer liquid sheet with \textit{We}\textsubscript{l} = 114 discharging from the gas-centered swirl coaxial atomizer CA2(\textit{S} =12.3) \textit{We}\textsubscript{l} remains constant for all the images shown in this figure.}
\end{figure*}
An image row in the figure corresponds to the cases of increasing Re$_g$ for a constant We$_l$ and an image column corresponds to the cases of increasing We$_l$ for a constant Re$_g$. It is evident from the image rows of Fig. \ref{Fig4} that for a given We$_l$, an increase in Re$_g$ destabilizes the outer liquid sheet and decreases the breakup length of the liquid sheet $L_b$. In a similar manner, an increase in We$_l$ for a given Re$_g$ increases $L_b$. Detailed analysis on the sheet breakup process was carried out by constructing image sequences for a particular $We_l$, however with varying Re$_g$. Figure \ref{Fig5} shows a sequence of images illustrating the breakup of the outer liquid sheet with $We_l =$ 114 under different central air jet flow conditions. 

The image at the leftmost top corner of Fig. \ref{Fig5} corresponds to the liquid sheet without any central air jet. The presence of the central air jet with low $Re_g$ pulls the liquid sheet toward the spray axis, as seen in the second and third images off of the top row of Fig.\ref{Fig5} and results in a shift in the sheet breakup location toward the orifice exit. The central air jet impinges directly on the inner surface of the liquid sheet at slightly higher Re$_g$ conditions, which may be leading to the generation of chunks of liquid masses immediately after the breakup. 
\begin{figure}[b!]
\centering\includegraphics[width=\linewidth]{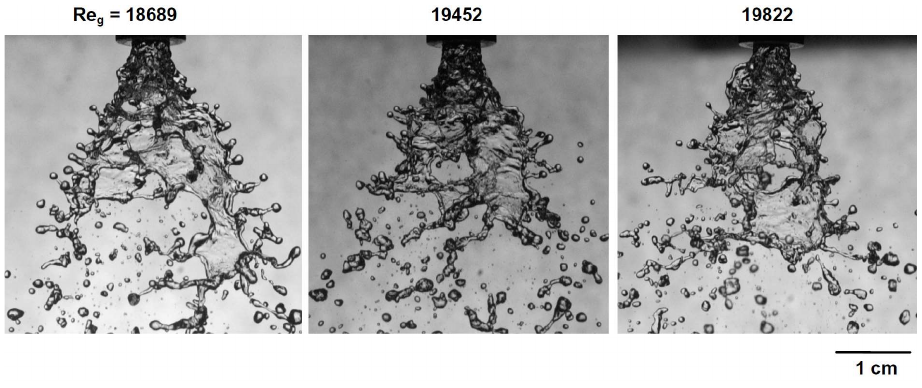}
\caption{\label{Fig6} High resolution images illustrating the violent and complex breakup of the liquid sheet by the central air jet in the near region of the orifice exit. Atomizer configuration is CA2 \textit{S} =12.3 and \textit{We}\textsubscript{l} = 114.}
\end{figure}
The liquid sheet collapses dramatically with further increase in Re$_g$, as seen in the images of Fig. \ref{Fig5} and the breakup of the liquid sheet occurs in the near region of the orifice exit. 
\begin{figure}[b!]
\centering\includegraphics[width=\linewidth]{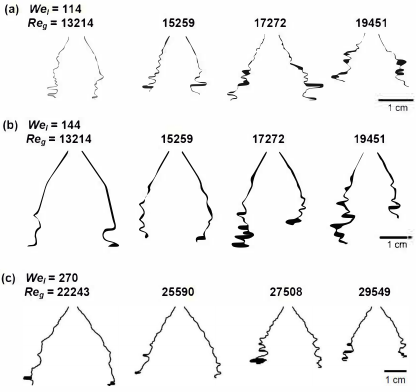}
\caption{\label{Fig7} Enlarged view of the two-dimensional surface profiles of liquid sheets discharging from the gas-centered swirl coaxial atomizer CA2 (\textit{S} =12.3) with different combinations of \textit{We}\textsubscript{l} and \textit{Re}\textsubscript{g}: (\textit{a}) \textit{We}\textsubscript{l} =114,(\textit{b}) \textit{We}\textsubscript{l} =144, and (\textit{c}) \textit{We}\textsubscript{l} =270}
\end{figure}
High resolution images illustrating the breakup of the liquid sheet at higher values of Re$_g$ are given in Fig. \ref{Fig6}. The axisymmetric conical shape of the outer liquid sheet, as observed in the lower values of Re$_g$, is no longer seen at these higher values of Re$_g$ and the sheet breakup occurs in a more complex manner. Discrete liquid sheets bounded by thick rims are dominantly seen at these flow conditions. The images given in Fig. \ref{Fig5} show an increase in the radial spread of spray with increasing Re$_g$ particularly in the higher range of $Re_g$ see the last five images of Fig. \ref{Fig5}. An intense mixing between the liquid sheet and the central air jet in the near region of the orifice exit occurs at these flow conditions because the jet boundaries are closer to each other than ever before. 

Note that the inertia of the central air jet in the near region of the orifice exit is proportional to  $\rho_gU_g^2$, which is two orders of magnitude larger than that of the liquid sheet, $\rho_lU_l^2$ at higher values of $Re_g$ given in Fig. 5. The increase in radial spread with increasing Re$_g$ at higher values of Re$_g$ may be a direct consequence of momentum exchange between the jets. A closer interaction between the liquid sheet and the central air jet in the near region of the orifice exit under these flow conditions may result in a violent breakup of liquid masses. 

Figures 7a-7c show the two-dimensional surface profiles of liquid sheets discharging from the gas-centered swirl coaxial atomizer CA2 for different combinations of We$_l$ and Re$_g$. In general surface corrugations seen on the liquid sheet increases with increasing Re$_g$ and this effect is more pronounced for low inertia liquid sheets. The liquid sheets with lower inertia are showing more surface corrugations for a given increase in $Re_g$, and higher inertia liquid sheets need a larger increase in Re$_g$ to develop similar levels of surface corrugations. Qualitatively this can be seen by comparing the profiles given in Fig. 7a or Fig. 7b with that of Fig. 7c. Note that the profiles of liquid sheets given in Fig. 7c correspond to higher values of Re$_g$ compared with those of Figs. 7a and 7b. The dark patches seen in a surface profile correspond to the presence of thick liquid masses, for example, liquid ligaments, and indicate the termination of a smooth liquid sheet. The profile analysis of liquid sheets with increasing Re$_g$ reveals that the dark patches appear more in the case of low inertia liquid sheets. This suggests that the low inertia liquid sheets are prone to develop thick liquid ligaments during the sheet breakup process. Interestingly it is observed that even at very high values of Reg, an intact liquid sheet is always present in the near region of the orifice exit and the sheet breakup occurs at distances one or two times the outer orifice diameter from the orifice exit. This can be understood from the sheet surface profiles given in Figs. 7a–7c. In the present gas-centered swirl coaxial atomizers, the liquid sheet at the orifice exit is separated from the central air jet by the inner orifice lip and a meaningful interaction between the liquid sheet and the central air jet is expected to happen a few millimeters away from the orifice exit.

\subsection{Quantitative Variation in $L_b$ With Flow Conditions}
\begin{figure}[b!]
\centering\includegraphics[scale=0.7]{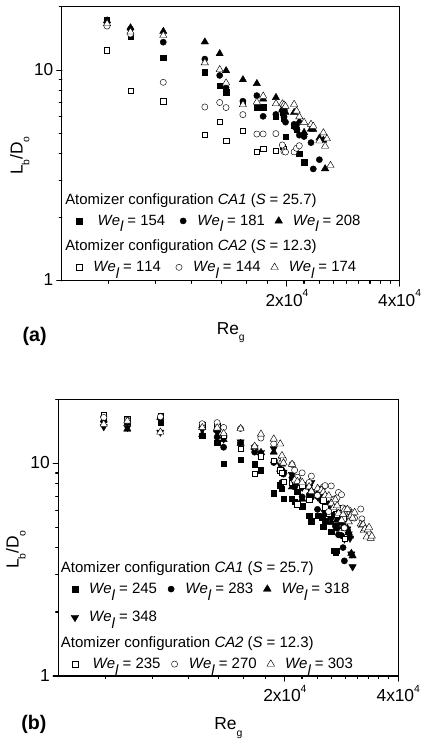}
\caption{\label{Fig8} Quantitative variation in \textit{L}\textsubscript{b} with \textit{Re}\textsubscript{g} for different values of \textit{We}\textsubscript{l} a low values of \textit{We}\textsubscript{l} and moderate and high values of \textit{We}\textsubscript{l}}
\end{figure}
The breakup process of outer liquid sheets is analyzed further by deducing quantitative measurements of $L_b$ from the spray images. Figures \ref{Fig8}a and \ref{Fig8}b show the quantitative variation in $L_b$ with Re$_g$ for different values of We$_l$. Each data point in Figs. \ref{Fig8}a and \ref{Fig8}8b corresponds to the arithmetic average of $L_b$ values measured from a set of a minimum of three images captured during repeated experimental runs at a given test condition. The variation in $L_b$ with Re$_g$ clearly shows that the central air jet destabilizes the outer liquid sheet. For low inertia liquid sheets, the influence of the central air jet on $L_b$ is seen even at low values of Re$_g$, as illustrated in Fig. \ref{Fig8}a. The variation in $L_b$ with Re$_g$ for high inertia liquid sheets given in Fig. \ref{Fig8}b shows that the breakup length remains unaltered in the lower values of Re$_g$ and drastic changes in $L_b$ are seen at higher Re$_g$. Figures  \ref{Fig8}a and \ref{Fig8}b also show that the decrease in $L_b$ with increasingRe$_g$ is more rapid for low inertia liquid sheets and is gradual for high inertia liquid sheets. The measurements of $L_b$ given in Figs. \ref{Fig8}a and \ref{Fig8}b do not show any significant influence of $S$ on the sheet breakup length for the atomizer configurations studied.
\subsection{Physical Insights on the Sheet Breakup Process}
\begin{figure}
\centering\includegraphics[scale=0.7]{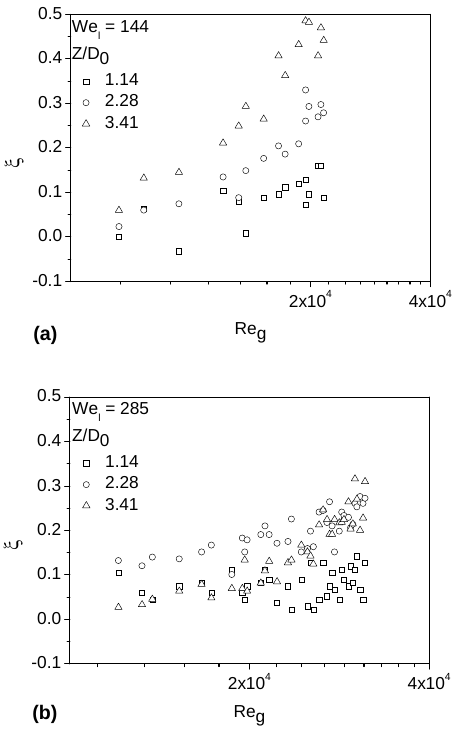}
\caption{\label{Fig9} Variation in spray contraction parameter with \textit{Re}\textsubscript{g} for different values of \textit{We}\textsubscript{l} at different axial locations, $Z$ from the orifice exit (\textit{a} and \textit{b}). Atomizer configuration is CA2, (\textit{S} = 12.3)}
\end{figure}

\begin{figure}[b!]
\centering\includegraphics[scale=0.7]{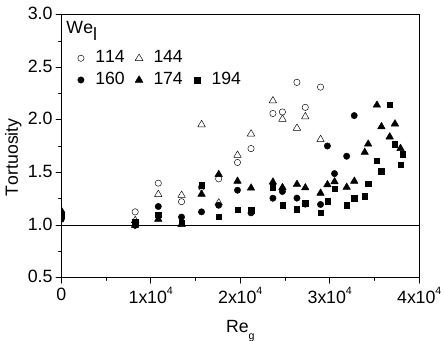}
\caption{\label{Fig10} Variation in tortuosity of the liquid sheet profile with Reg for liquid sheets with different \textit{We}\textsubscript{l} discharging from the gas-centered swirl coaxial atomizer CA2 (\textit{S} = 12.3)}
\end{figure}

The breakup process of liquid sheets in gas-centered swirl coaxial atomizers is influenced by several physical events. Since a swirling liquid sheet diverges away from the spray axis, it is essential to bring the liquid sheet closer to the boundary of the central air jet in order to establish an effective interaction process between the liquid sheet and the central air jet. This process is accomplished by the central air jet itself. The presence of the air jet inside the core of the outer swirling liquid sheet establishes an entrainment process of air between the central air jet and the inner surface of the outer liquid sheet. The air entrainment process reduces the local pressure over the inner surface of the outer liquid sheet and thereby increases the pressure difference across the liquid sheet $(P_e-P_i)$, where $P_e$ and $P_i$ are, respectively, the pressures at the outside and inside surfaces of the outer liquid sheet. For a given $P_e$, an increase in $(P_e-P_i)$ makes the liquid sheet move toward the spray axis. A stronger air entrainment process, or higher Re$_g$, increases $(P_e-P_i)$ and the spray continues to contract with increasing Re$_g$. Such a spray contraction with increasing Re$_g$ can be seen from the spray photographs given in Fig. \ref{Fig4}. In the present work a nondimensionalized spray contraction parameter is used to illustrate the trends of spray contraction with Reg and is expressed as,
\begin{equation}
\xi = \dfrac{SW_{Re_g=0} - SW_{Re_g}}{SW_{Re_g=0}}
\end{equation}
where $SW_{Re_g=0}$ and $SW_{Re_g}$ correspond to the spray widths for liquid sheets with Re$_g=0$ and Re$_g\neq0$, respectively. The variation in with Re$_g$ for the liquid sheets with different We$_l$ is shown in Figs. \ref{Fig10}a and \ref{Fig10}b. Since the liquid sheet is conical in nature, the variation in with Re$_g$ is presented in the figure for different axial distances $Z$ from the orifice exit. The measurements given in Fig. \ref{Fig9}a clearly suggest that the liquid sheets with lower We$_l$ exhibit a sharp increase in the spray contraction with increasing Re$_g$ compared with higher inertia liquid sheets. In other words, low inertia liquid sheets are more vulnerable to the presence of the central air jet and high inertia liquid sheets do not recognize the presence of the central air jet in the lower values of Re$_g$. 
Such a sharp increase in the spray contraction helps the liquid sheet to initiate a more active interaction with the central air jet. The second physical event to influence the sheet breakup process is the development of surface corrugations, which aid the generation of liquid ligaments from the sheet surface. Quantitative characterization of surface corrugations is done by estimating the tortuosity of the liquid sheet profile. Tortuosity of the liquid sheet profile at any two points in the profile is defined as the ratio of the length of the curved line profile between the two points to the least distance between the two points and is a measure of corrugations on the sheet profile exist between the two points. The value of tortuosity is always greater than 1 and increases with increasing levels of sheet profile corrugations. Assuming that the digitized points of the sheet profile are given in the \textit{x}-\textit{y} coordinate system as (\textit{x}\textsubscript{1} , \textit{y}\textsubscript{1}), (\textit{x}\textsubscript{2} , \textit{y}\textsubscript{2}), . . . , (\textit{x}\textsubscript{\textit{k}} , \textit{y}\textsubscript{\textit{k}}) , . . . , and (\textit{x}\textsubscript{\textit{n}} , \textit{y}\textsubscript{\textit{n}}), where subscripts 1 and \textit{n} correspond to the first and last points of the sheet profile curve, the tortuosity of the liquid sheet profile can be estimated as,
\begin{equation}
\textrm{Tortuosity} = \dfrac{\sum\limits_{k=1}^{n-1}{\sqrt{(x_{k+1} -x_k)^2 + (y_{k+1} -y_k)^2}}}{\sqrt{(x_n -x_1)^2 + (y_n -y_1)^2}}
\end{equation}
Attention must be given to ensure that the numerator was accurate enough to represent the total length of the curve. Figure \ref{Fig10} shows the variation in tortuosity of the liquid sheet profile with Re$_g$ for different values of $We_l$. The variation in tortuosity is almost insignificant for the liquid sheets with very low values of Re$_g$ 10,000–15,000, as seen in Fig. \ref{Fig10}.
A steep increase in the tortuosity with Re$_g$ is observed for low inertia liquid sheets in the moderate values of Re$_g$, as highlighted by the measurements marked with open symbols in Fig. \ref{Fig10} and the rise in tortuosity is somewhat gradual for relatively high inertia liquid sheets see the measurements marked with filled symbols in Fig. \ref{Fig10}. The third physical event to influence the sheet breakup process is the direct impingement of the central air jet on the inner surface of the liquid sheet and the subsequent mixing process between the liquid sheet and the central air jet. This event starts once the outer liquid sheet meets the boundary of the central air jet and generally occurs at high values of Re$_g$. The dynamics of the liquid sheet at these flow conditions is very complicated and is marked by features such as substantial reduction in the liquid sheet breakup length, generation of thick liquid ligaments and droplet clusters, cellular patterns on the surface of the liquid sheet, ejection of ligaments and droplets in the transverse direction, etc. The images shown in Fig. \ref{Fig6} are typical of this event. More systematic experiments are needed to ascertain these facts in a quantitative manner. The quantitative variation in $L_b$ with Re$_g$ given in Figs. \ref{Fig8}a and \ref{Fig8}b shows that the breakup process of liquid sheets discharging from gas-centered swirl coaxial atomizers is primarily determined by the flow parameters We$_l$ and Re$_g$. The influence of Re$_g$ on the sheet breakup is very significant for low inertia liquid sheets, whereas such behavior is observed only at higher values of Re$_g$ for high inertia liquid sheets. It is interesting to observe that low inertia liquid sheets disintegrate at lower values of Re$_g$. Note that low inertia liquid sheets discharging from a simple pressure swirl atomizer exhibit a longer breakup length and it is quite normal to presume that these liquid sheets require more energy for their breakup and the subsequent atomization process. 

A reverse trend is observed in the gas-centered swirl coaxial atomizers. This behavior is attributed to the fact that low inertia liquid sheets are more vulnerable to the entrainment process developed between the central air jet and the liquid sheet. High inertia liquid sheets are relatively less influenced by the entrainment process, which resulted in almost an insignificant variation in $L_b$ with Re$_g$ in the lower values of Re$_g$, as seen in Fig.  \ref{Fig8}b. Analysis of spray images revealed that the breakup of the liquid sheet is faster if the liquid sheet exists near the spray axis. This was observed for all values of $We_l$. Thus the primary trigger for the sheet breakup may be the intensity of entrainment process developed between the liquid sheet and the central air jet.

\section{Conclusions}
An experimental study on the breakup behavior of the outer swirling liquid water sheet by the central gas air jet discharging
from custom fabricated gas-centered swirl coaxial atomizers has been reported. The analysis of spray pictures captured using
conventional photographic techniques reveals that the presence of the central air jet in the core of the swirling liquid sheet significantly changes the breakup behavior of the liquid sheet. At low values of the air jet Reynolds number, the interaction process
between the liquid sheet and the central air jet develops corrugations on the surface of the liquid sheet. For a given low inertia
liquid sheet, the level of sheet corrugations, described in the present work in terms of tortuosity of the sheet profile, increases
with an increasing air jet Reynolds number. A reduced effect of the central air jet on the surface corrugations is observed for high
inertia liquid sheets. The breakup region of the liquid sheet is marked by features such as ejection of liquid ligaments from the
sheet surface, localized droplet clusters, increased surface corrugations on the liquid sheet, and cellular structures thin membrane
of the liquid sheet surrounded by a thick rim on the liquid sheet. Quantitative measurements of the liquid sheet breakup length suggest that the breakup length of low inertia liquid sheets decreases with an increasing air jet Reynolds number and the decrease in sheet breakup length is less severe for high inertia liquid sheets. The present flow configuration of liquid and air jets creates an air entrainment process between the inner surface of the liquid sheet and the central air jet. The entrainment process helps to bring the liquid sheet and the central air jet closer to each other, and thereby develops a more active interaction between the jets. A high inertia central air jet aids to develop a stronger air entrainment process and hence brings the liquid sheet nearer to the spray axis. This results in the direct impingement of the air jet on the liquid sheet and the breakup length of the liquid sheet decreases steeply at high inertia air jet conditions.

\section*{Acknowledgment} %% ASME requests this exact spelling, singular.
This work was supported by Space Technology Cell, Indian Institute of Science under Grant No. ISTC/MAE/SK/179.

\begin{nomenclature}
% capital letter comes first, lower case second
% don't capitalize first word of the definition
\entry{$D$}{diameter (m)}
\entry{$L$}{length (m)}
\entry{$t$}{thickness (m)}
\entry{$U$}{velocity (ms$^{-1}$)}

\EntryHeading{Greek Letters}
\entry{$\rho$}{density (kgm$^{-3}$)}
\entry{$\mu$}{dynamic viscosity (Pas)}
\entry{$\xi$}{spray contraction parameter (-)}
\entry{$\sigma$}{surface tension (Nm$^{-1}$)}

\EntryHeading{Dimensionless Groups}
\entry{Re}{Reynolds number, $\rho U D/\mu$}
\entry{We}{Weber number, $\rho U^2 D/\sigma$}

\EntryHeading{Superscripts and Subscripts}
\entry{b}{breakup}
\entry{i}{inner}
\entry{f}{liquid sheet}
\entry{g}{gas}
\entry{l}{liquid}
\entry{o}{outer}

\end{nomenclature}

%%%%%%%%%%%%%%%  APPENDICES  %%%%%%%%%%%%%%%%%%%%%%%%%%%%%%%%%%%%%%%%%

%% Note that appendices will be "numbered" A, B, C, ... etc. Use \section, not \section*
%% Subsections need not be numbered, use \subsection*
%% The equation counter is reset for each appendix
%% Figures will be numbered consecutively

\appendix   %%% starting appendices
\selectlanguage{english} 

% When you drop the [french] option, delete your .aux file as well, since [french] sets ":" as an active character.

%%%%%%%%%%%%%  BIBLIOGRAPHY  %%%%%%%%%%%%%%%%%%%%%%%%%%%%%%%%%%%%%%%%%

\nocite{*} %% <=== Delete this line - unless you wish to typeset the entire contents of your .bib file.

\bibliographystyle{asmejour}   %% .bst file that follows ASME journal format. Do not change.

\bibliography{JFE_Kulkarni_et_al_2010} %% <=== change this to name of your bib file

%%%%%%%%%%%%%%%%%%%%%%%%%%%%%%%%%%%%%%%%%%%%%%%%%%%%%%%%%%%%%%%%%%%%%%

%% To omit final list of figures and tables, use the class option [nolists]

\end{document}